\documentclass[printer]{aa}
\usepackage{graphicx}
\usepackage[varg]{txfonts}
\usepackage{color}

\begin{document}

\title{Comparison of solar radio and EUV synoptic limb charts during the present solar maximum}

\authorrunning{Oliveira e Silva et al.}
\titlerunning{Radio and EUV synoptic limb charts}

\author{A.~J.~Oliveira~e~Silva\inst{\ref{inst1}}\and C.~L.~Selhorst\inst{\ref{inst1},\ref{inst2}}$^\star$\and P.~J.~A.~Sim\~oes\inst{\ref{inst3}}\and C.~G.~Gim\'enez~de~Castro\inst{\ref{inst4},\ref{inst5}}}

\institute{
IP\&D - Universidade do Vale do Para\'iba - UNIVAP, S\~ao Jos\'e dos Campos, SP, Brazil ($\star$\email{caiuslucius@gmail.com})\label{inst1}
\and NAT - N\'ucleo de Astrof\'isica Te\'orica, Universidade Cruzeiro do Sul, S\~ao Paulo, SP, Brazil \label{inst2}
\and SUPA School of Physics and Astronomy, University of Glasgow, G12 8QQ, UK\label{inst3}
\and CRAAM, Universidade Presbiteriana Mackenzie, S\~ao Paulo, Brazil \label{inst4}
\and IAFE, Universidad de Buenos Aires/CONICET, Buenos Aires, Argentina\label{inst5}
}

\date{Received /
Accepted }

\abstract {}
{The present solar cycle is particular in many aspects: it had a delayed rising phase, it is the weakest of the last 100 years, and it presents two peaks separated by more than one year. To understand the impact of these characteristics on the solar chromosphere and coronal dynamics, images from a wide wavelength range are needed. In this work we use the 17~GHz radio continuum, formed in the upper chromosphere and the extreme ultraviolet (EUV) lines 304 and 171~{\AA}, that come from the transition region (He II, $T\sim 6-8\times10^4$~K) and the corona (Fe IX, X, $T\sim10^6$~K), respectively. { We here extend upon a }previous similar analysis, and compare the mean equatorial and polar brightening behavior at radio and EUV wavelengths during the maximum of the present solar cycle, covering the period between 2010 and 2015.}
{We analyze daily images at 304 and 171~{\AA} obtained by the Atmospheric Imaging Assembly (AIA) aboard the Solar Dynamics Observatory (SDO). The 17~GHz maps were obtained by the Nobeyama Radioheliograph (NoRH). To construct synoptic limb charts, we calculated the mean emission of delimited limb areas with 100" wide and angular separation of $5^\circ$.  }
{At the equatorial region, the results show an hemispheric asymmetry of the solar activity. The northern hemisphere dominance is coincident  with the first sunspot number peak, whereas the second peak occurs concurrently with the  increase in the activity at the south. The polar emission reflects the presence of coronal holes at both EUV wavelengths, moreover, the 17~GHz  polar brightenings can be associated with the coronal holes. Until 2013, both EUV coronal holes and radio polar brightenings were more predominant at the south pole. { Since then they have not been apparent} in the north, but thus appear in the beginning of 2015 in the south as observed in the synoptic charts.}
{ 
This work strengthens the association between coronal holes and the 17~GHz polar brightenings as it is evident in the synoptic limb charts, in agreement with previous case study papers. The enhancement of the radio brightness in coronal holes is explained by the presence of bright patches closely associated with the presence of intense unipolar magnetic fields. However, observations with better spatial resolution and also at different radio wavelengths will be necessary to fully understand the physical mechanisms that link these features.
}
 

\maketitle 

\section{Introduction}

Coronal holes are low density regions caused by the presence of locally open magnetic field lines, often observed at the solar poles. Due to the low density they are observed as dark regions in coronal EUV lines and X-rays. This characteristic is also observed at low radio frequencies formed in the corona. However, at frequencies higher than 15~GHz bright regions inside the the coronal holes are often observed \cite[see for example: ][]{Gopal1999,Brajsa2007}.

The presence of  bright polar regions is clear in 17~GHz maps, obtained by the Nobeyama Radioheliograph \citep[NoRH, ][]{Nakajima1994}, and shows a cycle mimicking the polar faculae cycle \citep{Selhorst2011,Gopal2012,Nitta2014}. \cite{Selhorst2010} proposed that these bright patches could be generated by small scale magnetic regions around the positions of intense magnetic patches (kG), that coincide with the polar faculae \citep{Tsuneta2008}. The presence of 17~GHz bright patches were also observed at lower latitudes and { have been }associated with the enhanced unipolar magnetic regions underlying the coronal holes \citep{Gopal1999,Maksimov2006}. On the contrary, small bright structures observed in EUV filter images have no one-to-one correspondence with the bright 17~GHz patches  \citep{Nindos1999,Riehokainen2001,Nitta2014}; and for this reason, \cite{Nitta2014} suggest that most of the latter  are artifacts created by the image synthesis and deconvolution used in NoRH maps.

\cite{Selhorst2010} compared the variation of the polar mean emission at radio (17~GHz) and EUV (304 and 171~{\AA}), obtained with the EIT/SOHO. The 304 and 171~{\AA} polar brightenings have distinct behaviors { during} the solar cycle: while 171~{\AA} is correlated with the solar cycle, 304~{\AA} is anti-correlated and in agreement with 17~GHz.
The authors addressed the 171~{\AA} variations to the presence of coronal holes, whereas the 304~{\AA} behavior was attributed to polar structures in the lower atmosphere, which also determine the polar emission at the 17~GHz radio continuum.

The formation of the He {\sc ii} 304 \AA\ emission depends on the conditions of the lower atmosphere (transition region and chromosphere) but can also be affected by direct photoionization produced by coronal EUV photons \citep{Hirayama:1971,Zirin:1975,Jordan:1975,Andretta:1997}. \cite{Worden:1999} compared the He {\sc ii} 304 \AA\ emission from different atmospheric structures, namely plage, enhanced network, active networks and quiet chromosphere, using EIT/SOHO images, over a period of two years during the solar minimum. They found that the intensity contrast of those structures (averaged over their areas) remains constant with time, when compared to the quiet chromosphere. The quiet chromosphere emission also does not vary during the period. We note however that their definition of quiet chromosphere also includes coronal holes. Worden et al. also suggest that due to the constency of the quiet chromosphere irradiance in the analyzed period, the He {\sc ii} 304 \AA\ long-term variations could be traced to the magnetic activity. This idea is supported by \cite{Pauluhn:2003}, who investigated magnetograms and chromospheric EUV irradiance during the period 1996--2000. They found that in quiet regions of the solar disk there was an increase of 10 to 20\% in the magnetic flux density, and a corresponding increase in irradiance in He {\sc i} 584 \AA, with similar characteristics to He {\sc ii} 304 \AA\ emission in the different atmospheric structures \citep[e.g. coronal holes, network regions][]{Glackin:1978}. We also note that \cite{Glackin:1978} indicate that both He {\sc i} 584 \AA\ and He {\sc ii} 304 \AA\ show limb brightening in quiet Sun regions and limb darkening in coronal holes. Finally, \cite{DelZanna:2011} showed that both He {\sc i} 584 \AA\ and He {\sc ii} 304 \AA\ are well correlated with the 10.7 cm flux over the period 1998--2010.

In this paper, we analyze the relationship between the limb emission at radio (17~GHz) and the EUV lines (304 and 171~{\AA}) using synoptic limb charts covering the period between 2010 and 2015. The EUV images were taken by the Atmospheric Imaging Assembly (AIA) \citep{Lemen2012} aboard the Solar Dynamics Observatory (SDO), launched in 2010, whereas the 17~GHz maps were obtained by the Nobeyama Radioheliograph. These data are described in the following section, with the results presented in Sect. 3. The discussion and main conclusions are presented in Sect. 4.

\section{Data Analysis}

\begin{figure*}
\centering
\includegraphics[width=17cm]{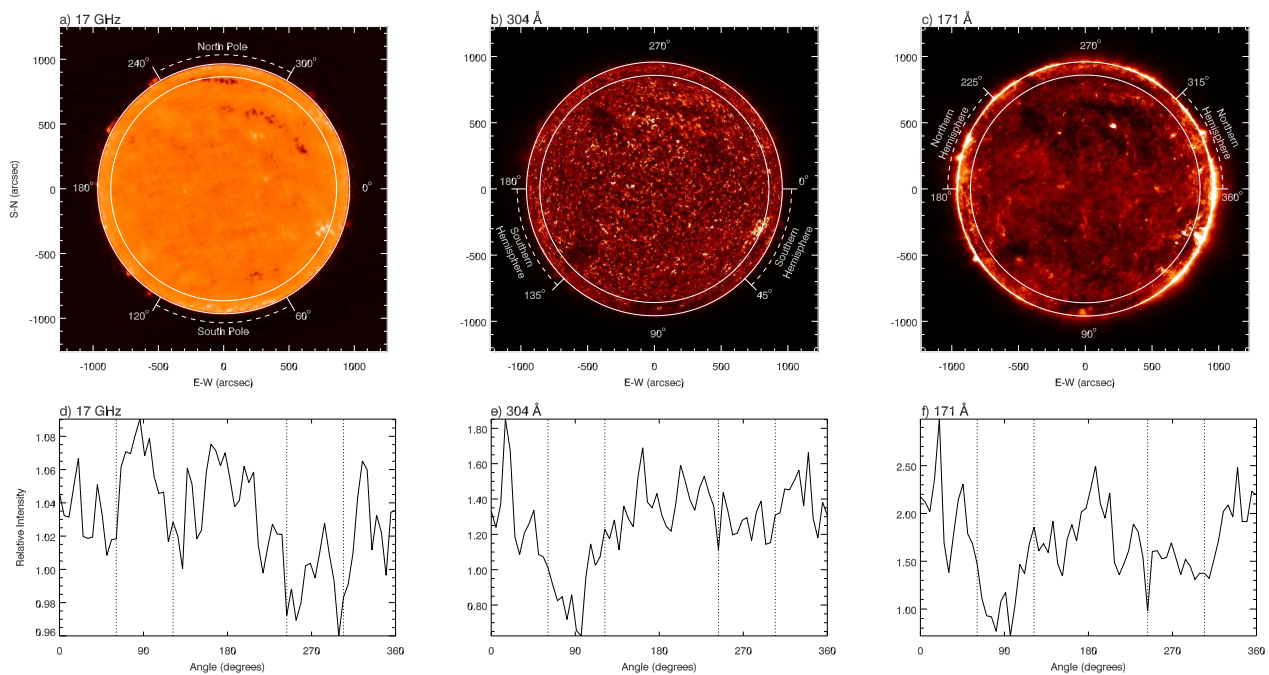} 
\caption{Solar images observed at (a) 17~GHz, (b) 304~{\AA}, and (c) 171~{\AA}. The concentric circles delimited the analyzed regions and the dashed arcs demarcated, respectively, the polar regions (a), the equatorial southern (b) and northern (c) hemispheres. The mean relative intensities ($I_R$) resulting of the analises for the three wavelengths are displayed in the bottom panels (d, e and f). The south and  north poles were delimited, respectively, in the ranges 60-120$^\circ$ and 240-300$^\circ$ (dotted lines in the panels).}
\label{fig:comp}
\end{figure*}

For the analysis presented here we use daily AIA EUV images taken with narrow band filters centered on He~{\sc ii} 304~\AA\ and Fe~{\sc ix} 171~\AA. While the 304~{\AA} line is an emission line formed in the transition region (TR), with temperatures ranging between 6 and $8\times10^4$~K, the 171~{\AA} comes from coronal heights, with temperatures of $T \sim 10^6$K. Images are $1024 \times 1024$ pixel size, and present an angular resolution of around 2".40. { The pre-reduced images were obtained at the Joint Science Operations Center (http://jsoc.stanford.edu/data/aia/synoptic/). To read the images, the standard AIA routines were used \citep[][]{Boerner:2012}, which are }part of the Solar SoftWare (SSW) package \citep{Freeland:1998}. { While, the 304~\AA\ images presented a constant exposure time of 2.90~s, at 171~\AA the first 68 days of the mission used exposure time of 2.90~s and 2.00~s after that. To make the 171~\AA data base uniform, the digital number (DN) of all maps were normalized  to  exposure time of 2.00~s.}

At 17~GHz (1.76~cm), solar maps have been obtained daily by the NoRH interferometer since 1992 with the same quality resolution (10--18").  { We analyzed the online daily pre-synthesized images (http://solar.nro.nao.ac.jp/norh/images/daily/), which are taken during local noon using a 10~s integration time. These maps uses the Steer CLEAN algorithm \citep{Steer1984}, that has the advantage of restoring diffuse features, with a ratio better than 1:300 between the RMS noise level and the peak brightness in the images \citep{Koshiishi2003}, which is very good for our work. The 17~GHz  maps} have $512 \times 512$ pixels and a spatial resolution of  $4.91$ arcsecs. Emission at 17~GHz mostly comes from the chromosphere (opacity $\tau=1$), where the local temperature and electron density are estimated to be $\sim 9.6\times10^3$~K  and $\sim 9.3\times10^9~{\rm cm}^{-3}$, respectively \citep{Selhorst2005a}. 

To investigate the relationship between the limb emission at radio and EUV wavelengths, we chose to construct synoptic limb charts. To produce these charts, we calculated the mean emission of delimited limb areas 100" wide and angular separation of $5^\circ$. The Fig.~\ref{fig:comp} shows examples of solar images obtained on May 14, 2010, at the three wavelengths studied here (respectively at 17~GHz, 304~{\AA} and 171~{\AA}) with the delimited limb region used to construct the synoptic limb maps. The angular positions were defined clockwise, with $0^\circ$ at the west limb, $90^\circ$ at the south pole, $180^\circ$ at the east and $270^\circ$ at the north pole. 


This procedure results in mean intensity profiles along the solar limb. To compare the variation of these profiles across the years, { we normalized the maps by the quiet Sun temperature (or Data Number (DN) for AIA images)}. The mean relative intensity ($I_R$) profiles for each map are displayed in the bottom panels of Fig.~\ref{fig:comp}. Distinct structures can be observed at the poles in the images: 

\begin{enumerate}
\item[i)] { The south pole} presents a coronal hole at 171~{\AA} (Fig.~\ref{fig:comp}c), that results in a noticeable reduction of the mean limb intensity between $60^\circ$ and $120^\circ$ (Fig.~\ref{fig:comp}f). The coronal hole and its associated mean intensity reduction are also observed at 304~{\AA} (Fig.~\ref{fig:comp}b and \ref{fig:comp}e). At 17~GHz we observe brights patches inside the EUV coronal hole area (Fig.~\ref{fig:comp}a and \ref{fig:comp}d) in agreement with previous observations \citep{Gopal1999,Maksimov2006}.  

\item[ii)]  { The north pole} also presents a dark area in the delimited region at 171~{\AA} (Fig.~\ref{fig:comp}c); however, it is a filament instead of a coronal hole, as can be seen in the 17~GHz map (Fig.~\ref{fig:comp}a). In this case, intensities at 17~GHz drop to values smaller than the quiet Sun level (Fig.~\ref{fig:comp}d). In the EUV images, the reduction in intensity is more prominent at 171~{\AA} (Fig.~\ref{fig:comp}f) than at 304~{\AA}, whose emission remains similar to the surrounding regions (Fig.~\ref{fig:comp}e).
\end{enumerate}

\begin{figure}[!h]
\centerline{ {\includegraphics[width=9cm]{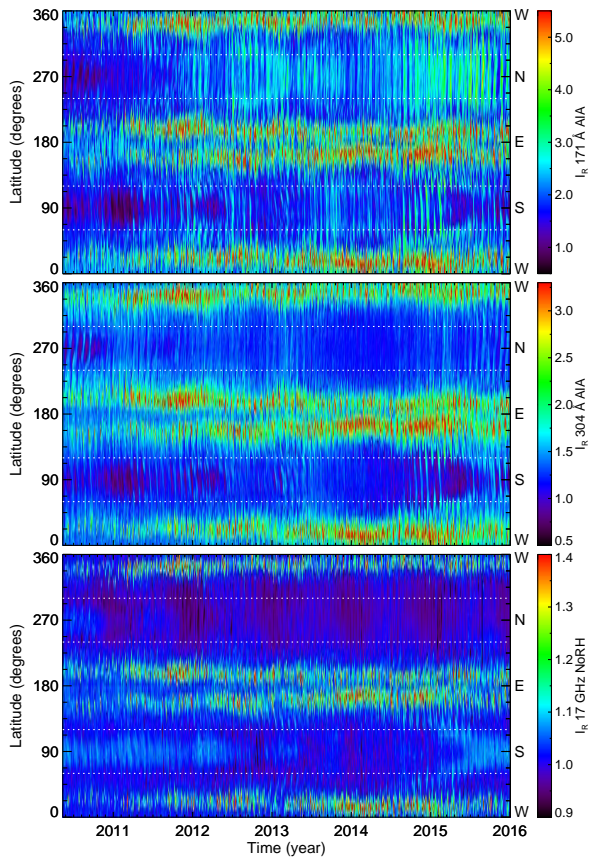}}}
\caption{Synoptic limb charts obtained for 171~\AA ~({\it top} panel), 304~{\AA} ({\it middle} panel) and 17~GHz ({\it bottom} panel). Running means of 3 days were applied in the synoptic charts of each wavelength. The minimum values of the color scales for the $I_R$ are 0.5 at the EUV (171 and 304~{\AA}) lines and 0.70 at radio (17~GHz), while the maximum values are 5.50, 3.30 and { 1.40} at 171~\AA, 304~{\AA} and 17~GHz, respectively. The dotted lines delimit the south ($60^\circ$ and $120^\circ$) and the north ($240^\circ$ and $300^\circ$) poles.  }
\label{fig:syn}
\end{figure}

{\section{The quiet Sun}

 The 17~GHz quiet Sun was determined for each individual map using a histogram ($bin=100$), in which the most common temperature was identified. Then we performed a weighed mean including two values immediately before the histogram { peak} and two values immediately after it, the result was assumed as the quiet Sun. Since the NoRH are normalized to $10^4$~K, the quiet Sun oscillates around it.

{ As reported by Boerner et al. (2014), the AIA responses have degraded over the years, most notably in the 304~{\AA} channel. Using the SolarSoft routine} {\it aia\_get\_response.pro}{, one can see the degradation of the AIA responses over time with respect to the EUV Variability Experiment (EVE) on board the SDO. At the end of 2015, the responses of the 171~{\AA} and 304~{\AA} channels decreased to $\sim$86\% and $\sim$20\%, respectively, of the initial values as of May 2010. We corrected the AIA 171~{\AA} and 304~{\AA} images for the time-dependent instrument degradation, and used them to calculate the quiet Sun intensity in DN with the same procedure that was used to obtain the 17 GHz quiet Sun brightness temperature from the NoRH images. The bin sizes were set to $\sim$1\% of the quiet Sun DN at the beginning of SDO observations, namely, 4 and 1 at 171~{\AA} and 304~{\AA}, respectively.

Figure~\ref{fig:quiet} shows the quiet Sun intensity derived in the above procedure. At 171~{\AA}, the quiet Sun presents an overall decrease over the years (black dots). The quiet Sun at 304~{\AA} (in grey) stayed almost flat until the latter half of 2014, when it started to decrease from $\sim$100 DN to $\sim$50 DN at the end 2015. The decrease of the quiet Sun in 304~{\AA} images since 2014 could represent true solar variations. However, it could be due also to residual calibration issues, as the last calibration of the AIA responses with the EVE took place in 2014, which is called version 6. The EVE calibration rocket scheduled for launch in 2016 will help us identify the origin of the downturn of the 304~{\AA} quiet Sun that started in 2014. We note that the synoptic results as presented in the next section and Figure~\ref{fig:syn} are not strongly affected by the apparent decrease of the quiet Sun intensity at 304~{\AA}.}

\begin{figure}[!h]
\centerline{ {\includegraphics[width=8cm]{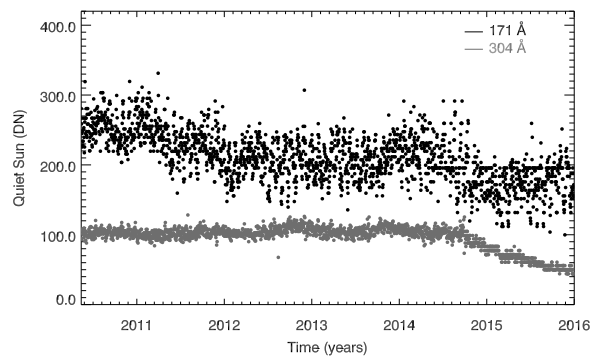}}}
\caption{The AIA quiet Sun variation at 171~{\AA} (black) and 304~{\AA} (grey). { The DN values are corrected for the time-dependent instrument responses.} }
\label{fig:quiet}
\end{figure}

\section{Results}

In the studied period (2010--2015), around 2000 images were analyzed for each wavelength, { after manually checking and discarding images with problems ($< 2\%$) }. The results at 17~GHz show that $I_R$ in the $5^\circ$ cells varies between 0.70 and 29.02; the smaller values could be addressed as the presence of dark filaments in the delimited region, and the higher values are due to the presence of {\it extreme} active regions, that can reach brightness temperatures above $5\times10^5$~K, i.e. 50 times greater than the quiet Sun values \citep[][and references therein]{Selhorst2014}. We attribute  the stronger emission from active regions to the  gyro-resonance contribution that becomes important in the presence of strong magnetic fields ($\sim 2000~G$) in the transition region or at coronal heights \citep{Shibasaki2011}. Nevertheless, these {\it extreme} active regions are rarely observed at the limb in the analyzed period: only 24 days present relative intensities  $I_R\geqslant5$ and 6 days have $I_R\geqslant10$.

In EUV filter images, the presence of coronal holes results in a stronger decrease of limb emission, with 304 and 171~{\AA} reaching $I_R$ minimum values of 0.57 and 0.49 respectively. On the other hand, the active regions at the limb increase $I_R$ at a lesser pace than in radio, with a maximum around 18.11 at 304~{\AA} and 26.42} at 171~{\AA}, { likely reflecting the single free-free emission mechanism that is not affected by strong magnetic field in active regions.} 

The daily $I_R$ angular distributions are organized in synoptic limb charts, in which the angular position is represented in the ordinates, and the time in the abscissas; intensities are color-coded. Since our main objective is to study the behavior of the polar regions as a whole, the $I_R$ values in Fig.~\ref{fig:syn} are artificially saturated at 5.50, 3.30 and { 1.40} at 171~{\AA}, 304~{\AA} and 17~GHz respectively in order to enhance the visualization of the tenuous variations at the polar regions. We now discuss the results in terms of the equatorial and polar regions separately.

\subsection{Equatorial emission}
\begin{figure}[!h]
\centerline{ {\includegraphics[width=8.cm]{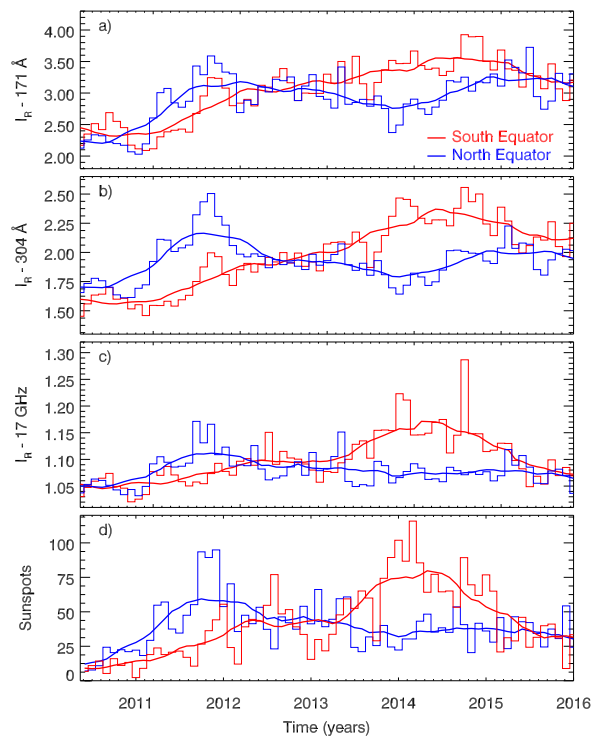}}}
\caption{Monthly mean equatorial $I_R$ at (a) 171~{\AA}, (b) 304~{\AA} and (c) 17~GHz. The red and blue curves are the mean equatorial $I_R$ in the south and north hemispheres, respectively. { For comparison, the monthly sunspot number for each hemisphere are also shown (d). Yearly running means, i.e., every 12 points, were applied (thick lines).}}
\label{fig:eq}
\end{figure}

In order to quantify the intensity increase in the equatorial region observed in Fig. \ref{fig:syn} and verify a possible north-south asymmetry in the solar activity, we calculated { the monthly mean emission} in the northern and southern hemispheres.  { The monthly mean} equatorial emission in the southern hemisphere is here defined as { the monthly average} of intensities in the ranges 0 to 45$^\circ$, and 135 and 180$^\circ$ (see Fig. \ref{fig:comp}b). In the northern hemisphere, it is defined as { the monthly average} between 180--225$^\circ$ and 315--360$^\circ$ (see Fig. \ref{fig:comp}c). The results are shown in Fig.~\ref{fig:eq}. For comparison with the solar cycle, the bottom panel shows { the monthly sunspot number (SSN) for each hemisphere (obtained at http://www.sidc.be/silso/datafiles). The southern and northern hemispheres are shown in red an blue respectively, and yearly running averages (12 points) were applied to all the values (thick lines).}

The equatorial results were are consistent with the hemispheric sunspot number. The  three wavelengths have similar mean equatorial intensity ($\langle I_R\rangle$) variations which is evidenct from the significant correlation coefficients between the radio and the EUV for the monthly averages (Table~\ref{table1}). Following the sunspot number, the synoptic charts showed a northern dominance until 2012 and the increase in the southern activity at the end of 2013. The differences between the equatorial hemispheric emission were more evident at the chromospheric level, at 304~{\AA} and 17~GHz (Fig.~\ref{fig:eq}b and \ref{fig:eq}c), which resulted in high correlation indexes, 0.79 and 0.80, for the southern and northern hemispheres respectively.

The hemispheric diferences at 171~{\AA} (Fig.~\ref{fig:eq}a ) are smaller than those observed in the chromosphere, however, the behavior is still the same. Moreover, good correlations were found in the comparison with the 17~GHz monthly values, namely 0.73 for the south and 0.62 for the north.





\subsection{Polar emission}

{ While in the equatorial analysis the EUV and radio emission showed a good agreement, at the poles the scenario changes completely and the differences between EUV and radio are clear.} At 171~{\AA} the presence of coronal holes, observed as dark regions at the poles (Fig.~\ref{fig:syn} {\it top}), are reflected in the synoptic map as a decrease in the mean intensity. The dark regions remain stable for months and are present in high number at the south pole. At the north pole the last large coronal hole was observed at the end of 2010 but, they can be observed in the southern hemisphere until the first half of 2012 { and after 2015}. Without the coronal holes both poles present an increase at 171~\AA. The coronal holes observed at 171~{\AA} were also identified in the 304~{\AA} synoptic chart (Fig.~\ref{fig:syn} {\it middle}).
On the other hand, at 17~GHz, instead of coronal holes the poles present a prominent intensity increase (Fig.~\ref{fig:syn} {\it bottom}), until 2012 { and after 2015}. In the absence of the coronal holes (around 2013), $I_R$ decreases.

\begin{figure}[!h]
\centerline{ {\includegraphics[width=8.cm]{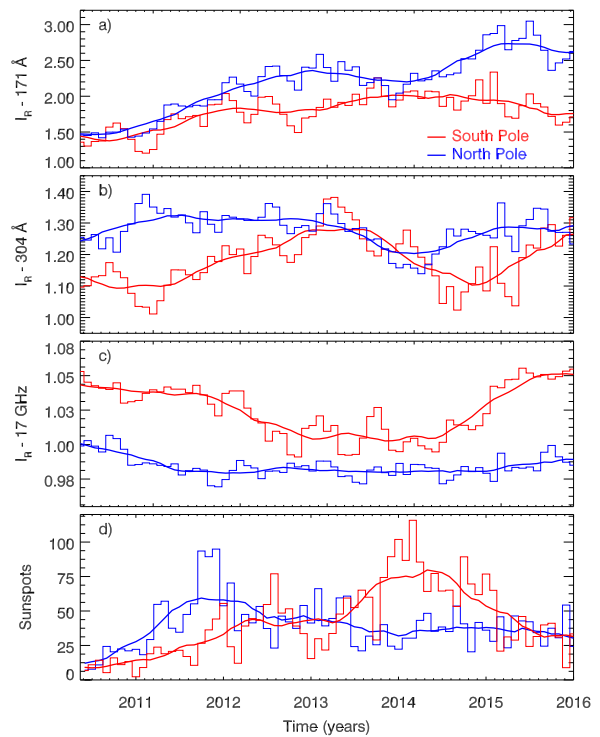}}}
\caption{Monthly mean polar $I_R$ variations at (a) 171~{\AA}, (b) 304~{\AA} and (c) 17~GHz. For comparison the monthly sunspot number for each hemisphere were plotted in panel (d). The daily south (red curves) and north (blue curves) means $I_R$ were calculated in the regions 60--120$^\circ$ and 240--300$^\circ$, respectively. { Yearly running means (12 points) were applied to the values (thick lines)}.     }
\label{fig:poles}
\end{figure}

In order to compare the temporal variation of the polar regions, the mean $\langle I_R\rangle$ of the south ($60^\circ$ and $120^\circ$) and north ($240^\circ$ and $300^\circ$) poles were calculated for each month. Results are shown in Fig.~\ref{fig:poles}, and present the following characteristics:

\begin{itemize}
\item {\em 171~{\AA}:} the mean $\langle I_R\rangle$ increases at both poles (Fig.~\ref{fig:poles}a) with the north (blue curve) more intense than the south pole (red curve); 
\item { {\em 304~{\AA}:} the mean $\langle I_R\rangle$ is almost constant at the north pole with a small reduction in the beginning of 2014 (Fig.~\ref{fig:poles}b, blue curves). In the south the mean $\langle I_R\rangle$ started to increase with the disappearance of the coronal holes and reached a maximum in 2013. Then $\langle I_R\rangle$ decreased to a minimum at the end of 2014 and, even with the presence of a coronal hole, it has started to increase again  (Fig.~\ref{fig:poles}b, red curves);}
\item { {\em 17 GHz:} contrarily to the EUV, the south pole presented greater mean $\langle I_R\rangle$ values during the analyzed period.  The north pole showed only small variations in the mean $\langle I_R\rangle$ during the analyzed period (Fig.~\ref{fig:poles}c, blue curves).  On the other hand, the south pole varied in the opposite way to the sunspot number, i.e.  the mean $\langle I_R\rangle$ decreased in the period in which the sunspot number reached the maximum (red curves in Fig.~\ref{fig:poles}c and \ref{fig:poles}d).}   
\end{itemize}

The correlation between the poles in radio and EUV resulted in smaller values than those obtained at the equator. In the south the indexes comparing 17~GHz with 171 and 304~{\AA} were, respectively, -0.39 and -0.36. These negative values can be attributed to the presence of coronal holes, which are dark at EUV, but presented a bright counterpart at 17~GHz. At the north pole, the coronal holes were less frequently observed, that resulted in a correlation index of -0.28, between 17~GHz and 171~{\AA} and a index close to zero (-0.04) in the comparison between 17~GHz and 304~{\AA}. 


\begin{table}
\caption{Correlation coefficients between radio and EUV results}
\begin{tabular}{lcccc}
\hline
\hline
& & & \multicolumn{2}{c}{ 17~GHz Correl. Coefficients}\\
\hline
& & & South & North\\
\hline
Equator &{\it vs } 171~{\AA}&   &  {0.73} & {0.62} \\
             &{\it vs } 304~{\AA}&  & {0.79} & {0.80} \\
\hline             
Pole      &{\it vs }171~{\AA} &  & ${-0.39}$ & ${-0.36}$ \\
             &{\it vs } 304~{\AA}&  & ${-0.28}$ & ${-0.04}$ \\
\hline
\end{tabular} 
\tablefoot{The correlation coefficients between the { monthly average of the} radio 17~GHz emission and the EUV emission at 171 and 304~{\AA}.
The analises were separated in the hemispheric mean polar and equatorial emission.}
\label{table1}
\end{table}

\section{Discussion and Conclusions}

We analyzed the polar and equatorial emission at 17~GHz, 171 ~{\AA} and 304~{\AA} through synoptic limb charts constructed from NoRH and AIA maps. The synoptic charts cover the period of { 2010-2015} show increasing solar activity in the equatorial region, as well as a reduction in the radio polar brightening { during the period of maximum activity}.

\subsection{Equatorial region}
At the equatorial region, the results for the three wavelengths clearly show the hemispheric asymmetry in the solar activity \citep[see for example:][]{Hathaway2010}. The dominance of the northern hemisphere coincided with the first SSN peak, whereas the second SSN peak occurs simultaneously with increase in the activity at the south (see Fig.~\ref{fig:eq}). There is an interval between the peaks (around the beginning of 2013), which indicates an inversion in the  asymmetry of the hemispheric activity. Moreover, the comparison between the hemispheres shows that the peak of activity in the south is more intense, { for all three wavelengths}.  

The stronger southern activity started in the first semester of 2013 and is coincident with the complete polar field reversion of the north pole \citep{Mordvinov2015}. Since solar cycle 20, polar field reversals occurred first in the north \citep{Svalgaard2013a}, all those cycles showed a more intense first peak. This trend has clearly changed in the present cycle, being the more intense the second one.   

The monthly equatorial radio $\langle I_R\rangle$ in both hemispheres presented an overall positive correlation with the EUV (see Table \ref{table1}). { The trend has clearly changed in the present cycle, in which the second peak is more intense.} A stronger activity in the southern hemisphere was more noticeable at 17~GHz and 304~{\AA}, both formed in the lower atmosphere.

\subsection{Polar regions}
In the polar regions, the presence of stable and long lived coronal holes is evident at both EUV wavelengths, identified by the darker patches in Fig.\ref{fig:syn}. We also note an asymmetry of their hemispheric activity, in which the darker patches at the north pole started to disappear earlier than at the south pole, in agreement with the results obtained by \cite{Karna2014}. { Remarkably, the south pole showed coronal holes in the beginning of 2015, whereas no coronal holes were observed at the north pole in the same period}. 

The 17~GHz radio emission at the south pole clearly { decreases while the SSN increases, and vice-versa} (Fig.~\ref{fig:poles}); on the other hand, the radio emission at the north pole is almost constant during the studied period. { For the EUV bands, $\langle I_R\rangle$ at the north pole is more intense than the $\langle I_R\rangle$ at the south throughout the entire analysed period. At 17 GHz, the south pole dominated in the period \citep[cf.][]{Selhorst2011,Gopal2012,Nitta2014}.}

{ This work strengthens the association between coronal holes and the 17~GHz polar brightenings as it is evident in the synoptic limb charts showed in Fig.~\ref{fig:syn}, in agreement with previous case study works \citep{Gopal1999,Maksimov2006}. The enhancement of the radio brightness in coronal holes is explained by the presence of bright patches closely associated with the presence of increased unipolar magnetic regions underlying the coronal holes \citep{Gopal1999,Brajsa2007,Selhorst2010}. However, the physical mechanisms that link the radio brightenings and coronal holes are not yet fully understood. Observations of these brightenings with better spatial resolution and also at different radio wavelengths, such as solar observations with the Atacama Large Millimetric/Submillimetric Array (ALMA) \citep{2015SSRv..tmp..118W} might be fundamental to investigate this association.}






\begin{acknowledgements}
 { The authors would like to thank the referee for their valuable comments and suggestions which helped to improve the manuscript. We also thank Peter Levens for language corrections.} We would like to thank the Nobeyama Radioheliograph, which is operated by the NAOJ/Nobeyama Solar Radio Observatory. A.J.O.S. acknowledge the scholarship form CAPES. C.L.S. acknowledge financial support from the S\~ao Paulo Research Foundation (FAPESP), grant number 2014/10489-0. 
\end{acknowledgements}


\bibliographystyle{aa}

\end{document}